# Spin-Hall Topological Hall Effect in Highly Tunable Pt/Ferrimagnetic-Insulator Bilayers

*Adam S. Ahmed[1,*], Aidan J. Lee[1,*], Nuria Bagués[2], Brendan A. McCullian[1], Ahmed M. A. Thabt[1], Avery Perrine[2], James R. Rowland[1], Mohit Randeria[1], P. Chris Hammel[1], David W. McComb[2,3], and Fengyuan Yang[1]*

[1]Department of Physics, The Ohio State University, Columbus, OH, USA

[2]Center for Electron Microscopy and Analysis, The Ohio State University, Columbus, OH, USA

[3]Department of Materials Science and Engineering, The Ohio State University, Columbus, OH, USA

[*]Equal Contributions

**Electrical detection of topological magnetic textures such as skyrmions is currently limited to conducting materials. While magnetic insulators offer key advantages for skyrmion technologies with high speed and low loss, they have not yet been explored electrically. Here, we report a prominent topological Hall effect in Pt/$Tm_3Fe_5O_{12}$ bilayers, where the pristine $Tm_3Fe_5O_{12}$ epitaxial films down to 1.25 unit cell thickness allow for tuning of topological Hall stability over a broad range from 200 to 465 K through atomic-scale thickness control. Although $Tm_3Fe_5O_{12}$ is insulating, we demonstrate the detection of topological magnetic textures through a novel phenomenon: "spin-Hall topological Hall effect" (SH-THE), where the interfacial spin-orbit torques allow spin-Hall-effect generated spins in Pt to experience the unique topology of the underlying skyrmions in $Tm_3Fe_5O_{12}$. This novel electrical detection phenomenon paves a new path for utilizing a large family of magnetic insulators in future skyrmion technologies.**

There have been intense efforts to find new platforms for next-generation magnetic memory with superior speed and lower energy consumption that surpass current technology. Magnetic skyrmions offer promise to meet these requirements owing to their unique topological protection—a property that allows a magnetic bit to exist with sizes less than 10 nm and maintain robust stability.[1-3] To date, perpendicularly-magnetized metallic ferromagnetic (FM) multilayers, such as [Pt/Co/Ir]$_N$, have shown room temperature skyrmions with sizes down to 30 nm.[4] However, these metallic polycrystalline multilayers typically possess a considerable amount of pinning sites from grain boundaries and relatively high damping loss, which will limit their read/write speeds and energy efficiency.[5] To this end, high-quality epitaxial magnetic heterostructures are desired for high-speed, low-loss, and broadly tunable skyrmion memory devices.

As compared with metallic FMs, ferrimagnetic insulators (FMI) avoid resistive losses and Joule heating due to their insulating nature. Meanwhile, the magnetic information can be read and controlled using a heavy-metal (HM)/FMI bilayer structure, where the spins generated by spin Hall effect (SHE) in the HM can couple strongly to the magnetization of the FMI at the interface, enabling high-efficiency manipulation of the FMI magnetization.[6] For example, FMIs such as $Y_3Fe_5O_{12}$ (YIG) and $Tm_3Fe_5O_{12}$ (TmIG) can be grown epitaxially with pristine quality, atomically sharp interfaces, low magnetic damping ($10^{-4}$–$10^{-2}$) allowing for low switching energies and tunable magnetic anisotropy from in-plane (IP) to out-of-plane (OOP) via epitaxial strain.[6-8] Despite these benefits, skyrmions have been predominantly observed in conducting materials, and rarely studied in insulators (e.g., bulk $Cu_2OSeO_3$).[9] As indicated by skyrmions in metallic multilayers, HM/FMI thin bilayers, in principle, should be able to form skyrmions enabled by a sizable interfacial Dzyaloshinskii-Moriya interaction (DMI).[10] More attractively, FMI epitaxial films enable strain engineering to control magnetic anisotropy and Curie temperature ($T_C$) through precise thickness control, resulting in powerful tunability of skyrmion properties.

In this article, we report the creation of topological magnetic textures in Pt/TmIG bilayers and detection of such textures in TmIG thin films using the Pt layer via a novel detection scheme which we coin "spin-Hall topological Hall effect" (SH-THE) arising from the spin-torque interactions at the Pt/TmIG interface. Our results demonstrate the highly tunable phase space of strong SH-THE over a broad temperature range of 200 – 465 K by precisely and reproducibly controlling the TmIG thickness down to the atomic level, paving the path for the design and implementation of skyrmion devices for operation at room temperature (RT) or other desired temperatures ($T$). The excellent crystalline quality and pristine interfaces of our TmIG films is verified by cross-sectional and plan-view scanning transmission electron microscopy (STEM), allowing for the detection of SH-THE in Pt/TmIG bilayers down to 1.5 nm (1.25 unit cells) TmIG.

Pt(3 nm)/TmIG($t$) bilayers with TmIG thickness ($t$) from 1.5 to 8 nm are grown on (111)-oriented $Gd_3Ga_5O_{12}$ (GGG) and substituted-GGG (sGGG) substrates with lattice constants of 12.382 Å and 12.480 Å, respectively, using ultrahigh vacuum off-axis sputtering.[11, 12] As a comparison, the bulk lattice constant of TmIG is 12.324 Å, resulting in a tensile strain of -0.47% on GGG and -1.25% on sGGG. X-ray diffraction (XRD) and STEM imaging are utilized to investigate the crystalline quality of the TmIG layer. The $2\theta$-$\omega$ XRD scans of 35 nm TmIG films grown on GGG and sGGG in Fig. 1a both exhibit clear Laue oscillations. The TmIG(444) peak at 51.299° and 51.722° corresponds to the film on GGG and sGGG with OOP lattice constants of 12.329 Å and 12.235 Å, respectively. The XRD rocking curves of the TmIG(444) peak in Figs. 1b and 1c give exceptionally narrow full-width-half-maxima (FWHM) of 0.0051° and 0.0060° for the 35 nm TmIG film on GGG and sGGG, respectively, reaching our instrument limit. The presence of pronounced Laue oscillations and sharp rocking curves demonstrate the high crystalline uniformity of the TmIG films on both substrates.

To visualize the crystalline ordering, we perform both cross-sectional and plan-view imaging of a Pt(5 nm)/TmIG(5 nm) bilayer on sGGG(111) using STEM. Figure 1d shows a cross-sectional STEM image of the bilayer viewed along the [1$\bar{1}$0] direction, where the yellow dashed line indicates the TmIG/sGGG interface with perfect continuation of the garnet lattice without any detectable dislocations or other defects throughout the thickness of the TmIG film. Figure 1e shows a plan-view STEM image exhibiting a remarkable snowflake pattern. The high-magnification image in Fig. 1f reveals fine details of the atomic arrangement of concentric hexagons, where a [111]-projection of the garnet lattice is overlaid on the STEM image to show a perfect match. The combination of XRD scans and STEM images confirms the state-of-the-art crystalline ordering throughout the whole film.

The static and dynamic magnetic properties of TmIG are characterized using a vibrating sample magnetometer (VSM) and a cavity ferromagnetic resonance (FMR) spectrometer. Figure 2a shows the magnetization ($M$) hysteresis loop in an OOP magnetic field ($H$) of a TmIG(5 nm)/sGGG film after subtraction of the paramagnetic background. The TmIG thin film exhibits a sharp magnetic reversal at a coercive field of $H_c$ = 1.3 Oe and a saturation magnetization $M_s$ = 86 emu/cm$^3$ which is slightly smaller than the bulk value of 110 emu/cm$^3$.[13] The extremely small coercivity of our TmIG films is to date the lowest reported with the sharpest switching, suggesting high uniformity and low defect density, which imply that very low energies are needed to switch the TmIG magnetization, magnetically or electrically.[6, 8, 14-16]

We measure the $T_C$ of a 60 nm TmIG film in a VSM oven with a 5 Oe OOP field as shown in Fig. 2b, from which $T_C$ = 508 K is determined. This $T_C$ is well above RT desired for practical applications, while it can be finely tuned to be near RT by reducing the film thickness for the stabilization of skyrmions.[17, 18] For TmIG films thinner than 5 nm, the magnetometry signals are too weak to be detected reliably; thus, we use cavity FMR at RT to measure very thin TmIG films down to 1.85 nm. Figure 2c shows the OOP FMR derivative spectra at 9.62 GHz for 1.85 nm TmIG films on GGG and sGGG, which can be fit to the derivative of Lorentzian function to extract the peak-to-peak linewidth ($\Delta H$) and resonance field ($H_{res}$). The linewidths for the films grown on GGG and sGGG are $\Delta H$ = 153 Oe and 166 Oe, respectively.[19, 20] The OOP resonance fields for the two samples are significantly different ($H_{res}$ = 5574 Oe on GGG and 3401 Oe on sGGG), indicating that the magnetic anisotropy vary greatly due to epitaxial strain.[7, 20]

The magnetic anisotropy of TmIG thin films can be determined from the angular dependence of FMR between OOP ($\theta$ = 0°) and IP ($\theta$ = 90°) as shown in Fig. 2d for the 1.85 nm TmIG films on GGG and sGGG, which exhibit opposite angular dependencies. This demonstrates that the TmIG film grown on GGG has in-plane anisotropy and that on sGGG has perpendicular magnetic anisotropy[19] due to the different tensile strain imparted on the TmIG film grown on the two substrates. The TmIG grown on GGG has a lattice mismatch of -0.47%, for which the strain-induced OOP magnetocrystalline anisotropy is not large enough to overcome the shape anisotropy, resulting in an in-plane anisotropy. On the contrary, TmIG on sGGG has a significantly larger lattice mismatch of -1.25%, which produces a stronger OOP magnetocrystalline anisotropy and makes the TmIG film perpendicularly magnetized. Therefore, growing TmIG films on these two substrates allows us to control the anisotropy for tuning the phase space of topological magnetic textures. In addition, Fig. 2d also confirms that the 1.85 nm TmIG is ferrimagnetic at RT. This very thin, OOP-magnetized FMI film combined with a capping layer with large spin-orbit coupling provides a unique platform for studying the emergence of nontrivial magnetic textures such as skyrmions in the presence of interfacial DMI.

Topological Hall effect is a major signature for detecting topological magnetic textures such as skyrmions. Because it is generally challenging to grow magnetic garnet films with sufficiently high quality for thicknesses below 5 nm, we first measure Hall resistivity for samples with 8 and 5 nm TmIG. However, we only detect the AHE-like signals without any sign of the THE at temperatures up to our instrument limit of 465 K, which suggests that the $T_C$ of the TmIG films may be too high. Therefore, we fabricate thinner TmIG films and discover highly tunable THE for TmIG thickness between 1.5 and 3.0 nm. Figure 3a shows the Hall resistivity ($\rho_{xy}$) for a Pt(3 nm)/TmIG(1.85 nm)/sGGG(111) sample from $T = 265$ to 365 K. A peak starts to emerge at 265 K and grows with increasing temperature, which after reaching a maximum at 285–300 K, eventually vanishes at 365 K.

In general, for the HM/FMI bilayers, the total Hall resistivity may include the following possible contributions,

$$\rho_{xy} = \rho_{OH} + \rho_{AH,Pt} + \rho_{SH-AH} + \rho_{TH,Pt} + \rho_{SH-TH}, \qquad (1)$$

where $\rho_{OH}$ is the ordinary Hall resistivity proportional to $H$ (subtracted from Fig. 3a), $\rho_{AH,Pt}$ is the anomalous Hall effect (AHE) term generated by the proximity-induced magnetization in Pt, $\rho_{SH-AH}$ is the spin-Hall anomalous Hall term[6, 19] arising from the spin torque interactions with the magnetization of TmIG at the Pt/TmIG interface, $\rho_{TH,Pt}$ is the topological Hall resistivity due to the proximity-induced magnetization in Pt, and $\rho_{SH-TH}$ is the spin-Hall topological Hall resistivity originating from the spin-torque interactions between the spin-Hall-effect produced spins in Pt and the topological magnetic textures in TmIG.

The last four terms in Eq. (1) include two from AHE ($\rho_{AH,Pt}$, $\rho_{SH-AH}$) and two from THE ($\rho_{TH,Pt}$, $\rho_{SH-TH}$), where $\rho_{AH,Pt}$ and $\rho_{TH,Pt}$ are caused by the proximity effect in Pt. As discussed below, there is no magnetic proximity effect in our Pt/TmIG systems. Therefore, the observed topological Hall resistivity, $\rho_{TH}$, is exclusively from $\rho_{SH-TH}$. In addition, the measured AHE resistivity solely comes from $\rho_{SH-AH}$ which needs to be subtracted to extract the topological Hall resistivity. The AHE resistivity should follow the magnetic hysteresis loop, for which the moment is too small to be measured reliably for the 1.85 nm TmIG film. Since the AHE resistivity cannot exceed the value at high field when the magnetization is saturated, we approximate the AHE contribution as a $\tanh(H/H_0)$ function, where $H_0$ is a fitting parameter (see Fig. 3b as an example for 300 K). By taking the difference between $\rho_{xy}$ and the AHE function, we obtain the topological Hall resistivity in Fig. 3c at RT, which reveals a prominent $\rho_{TH}$ peak located at ~800 Oe with a TH resistivity of 3.9 n$\Omega$ cm and a broad field range of ~4 kOe for RT topological Hall stability. We point out that the AHE function in Fig. 3b may not accurately describe how rapidly the magnetization is reversed in our films. For example, the reversal of magnetization may take place within a wider field range than the $\tanh(H/H_0)$ function, and as a result, the magnitude of $\rho_{TH}$ presented here is a lower bound.

After observing the topological Hall effect in Pt/TmIG(1.85 nm)/sGGG(111) at room temperature, we perform similar measurements on bilayers with TmIG thicknesses of 1.5, 2.0, and 3.0 nm, and detect THE at quite different temperature regimes. Figure 4 show the *H-T* diagrams of the topological Hall resistivity for these bilayers to visualize the phase space of $\rho_{TH}$, where Fig. 4b corresponds to the Pt/TmIG(1.85 nm)/sGGG data shown in Fig. 3. Immediately, we notice a systematic trend with increasing TmIG thickness: the region of topological Hall stability shifts towards higher temperatures as the TmIG thickness increases. The THE phase space occupies 210-290 K for the 1.5 nm TmIG, and rises to 265-355 K, 300-380 K, and 405-465 K for the 1.85, 2.0, and 3.0 nm samples, respectively. For the 5 nm TmIG film (Fig. 4e), $\rho_{TH} = 0$ up to 465 K. Thus, we show remarkable tunability of the THE with precise thickness control. Furthermore, we believe the temperature scaling with thickness is proportional to the $T_C$ which generally decreases with reducing dimensionality.[17, 18] This region of topological Hall stability is similar to the B20 skyrmion materials, where the presence of skyrmions and THE typically appear in a temperature window just below $T_C$.[21-23]

It has been predicted that magnetic anisotropy can be a tuning parameter for topological magnetic domains and skyrmions.[24, 25] Given our ability to control magnetic anisotropy by epitaxial strain, we check the influence of magnetic anisotropy by performing Hall measurements on a Pt(3 nm)/TmIG(2 nm)/GGG(111) where the TmIG film has in-plane anisotropy (see Fig. 2d). By comparing Figs. 4c and 4f, we note that the region of topological Hall stability is qualitatively similar in *M-H* space. Thus, magnetic anisotropy is not a dominant tuning parameter for the topological Hall phase in our films.

Lastly, we discuss the origin of the novel electrical detection of THE in an HM adjacent to an FMI. Equation (1) includes two anomalous Hall terms. The conventional AHE contribution ($\rho_{AH,Pt}$) is finite if the Pt layer is magnetic due to the proximity effect. The spin-Hall anomalous Hall term ($\rho_{SH-AH}$) arises from spin-orbit torques between the spin-Hall-effect induced spin polarization in Pt and the magnetization in TmIG. Two other terms, $\rho_{TH,Pt}$ and $\rho_{SH-TH}$, may exist for the topological Hall contributions. Since two of the four terms mentioned above originate from the proximity effect in Pt, it is important to verify or rule out the presence of the magnetic proximity effect. We perform anisotropic magnetoresistance (AMR) measurements at room temperature for the Pt(3 nm)/TmIG(1.85 nm)/sGGG sample (where we observe a large $\rho_{TH}$; see Fig. 3c), which only exhibits ordinary magnetoresistance (see Supplementary Information). This control experiment proves that the Pt layer is not magnetic and $\rho_{AH,Pt} = \rho_{TH,Pt} = 0$, which leads to $\rho_{xy} = \rho_{OH} + \rho_{SH-AH} + \rho_{SH-TH}$. Therefore, we conclude that the observed $\rho_{TH}$ signal comes from a novel "spin-Hall topological Hall effect" detection scheme, where the Pt spins sense the emergent electromagnetic field arising from the topological magnetic textures in TmIG through interfacial spin-torque interactions.[10] Additionally, since the damping-like torque dominates at the Pt/TmIG interface,[15] the Pt spins should follow the wrapping of magnetic textures and acquire a non-vanishing Berry phase. However, the exact nature of the interactions at the interface requires future investigation.

In conclusion, we have shown fine tuning of topological Hall stability by precise control of the TmIG thickness in Pt/TmIG bilayers. This allows us to predict and stabilize topological magnetic textures, like skyrmions, over a broad range of temperatures, including below, at, and above room temperature—a powerful tuning knob for integrating and advancing magnetic memory technology. These desired capabilities are enabled by the high-quality epitaxial films with fine control over the magnetic properties. Finally, we have discovered a new electrical detection scheme—spin-Hall topological Hall effect—which will now enable magnetic insulators to be viable materials platforms for topologically-driven technologies.

*Note added*: Recently we become aware of an independent report[26] which also detected topological Hall effect in HM/FMI systems.

**Acknowledgements**

This work was primarily supported by DARPA under Grant No. D18AP00008. Partial support was provided by the Center for Emergent Materials, an NSF-funded MRSEC, under Grant No. DMR-1420451 for the bulk synthesis and film growth of TmIG.

**Methods**:

**Sample growth and preparation:** The TmIG sputtering target was synthesized using a sol-gel method, which yielded a pure phase TmIG powder that was pressed into a two-inch sputtering target. The TmIG epitaxial films were grown using ultrahigh vacuum off-axis sputtering at a substrate temperature of 700°C on (111)-oriented GGG and sGGG substrates. The substrates were then cooled down to room temperature before growing the Pt layer.

The sample imaged in the cross-sectional orientation was prepared using a FEI Helios NanoLab 600 DualBeam Focused Ion Beam with Ga ion milling and polishing performed at 30 and 5 kV, respectively. The plan-view sample was made via wedge mechanical polishing using an Allied Multi-prep that resulted in a wedge sample with a thickness less than 20 $\mu$m with a 2-degree slope. High angle annular dark field (HAADF) STEM was performed on an image aberration corrected FEI Titan 60/300 STEM operated at 300 kV.

**Magnetotransport measurements:**

Standard photo-lithography and Ar ion milling were used to fabricate 100 $\mu$m wide Hall bar patterns on Pt(3 nm)/TmIG($t$) samples grown on GGG and sGGG. Hall measurements were carried out in a cryostat and oven in an electromagnet with a range of 1.6 T and a maximum temperature of 465 K. A source current of 300 $\mu$A was applied to the sample that results in a current density of $1 \times 10^{-9}$ A m$^{-2}$.

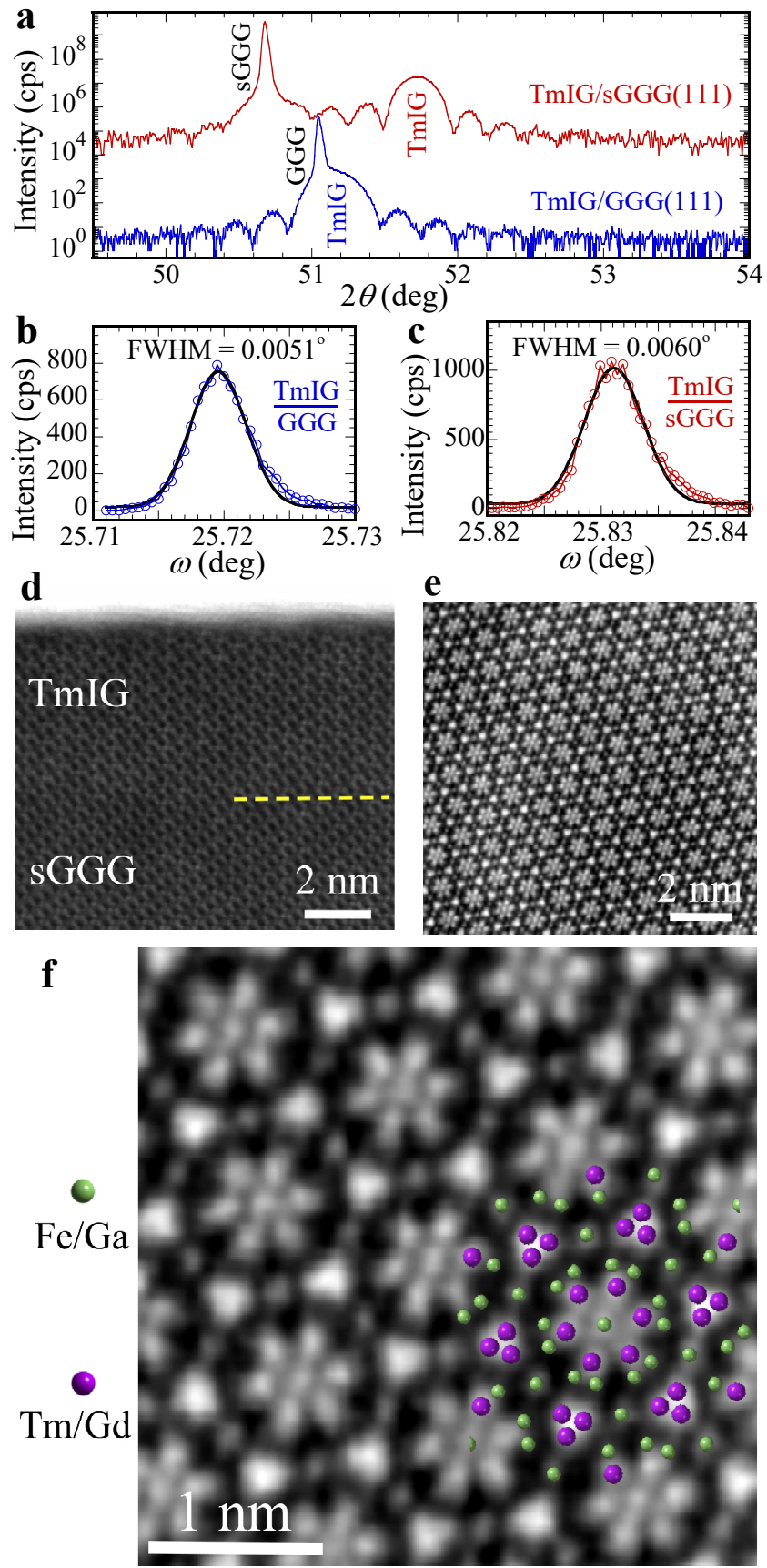

**Figure 1**. **Structural characterization of TmIG films. a**, XRD 2$\theta$-$\omega$ scans of TmIG(35 nm) grown on (111)-oriented GGG (blue) and sGGG (red) resulting in different tensile strain. XRD rocking curves of the TmIG(444) peak grown on **b**, GGG and **c**, sGGG. **d**, Cross-sectional STEM images for a Pt(5 nm)/TmIG(5 nm)/sGGG(111) sample viewed along the [1$\bar{1}$0] direction. The dashed yellow line marks TmIG/sGGG interface. **e**, Plan-view STEM image of a TmIG(5 nm)/sGGG(111) sample. **f**, High-magnification plan-view STEM image with an overlay of the [111]-projection of garnet lattice to highlight the snowflake pattern. Each bright corner of a hexagon is a closely-packed three Tm/Gd columns in a triangular arrangement. Inside the bright hexagon is another (smaller) hexagon of Tm/Gd columns. Fe/Ga columns occupies the center of the two concentric hexagons and decorate between the Tm/Gd columns.

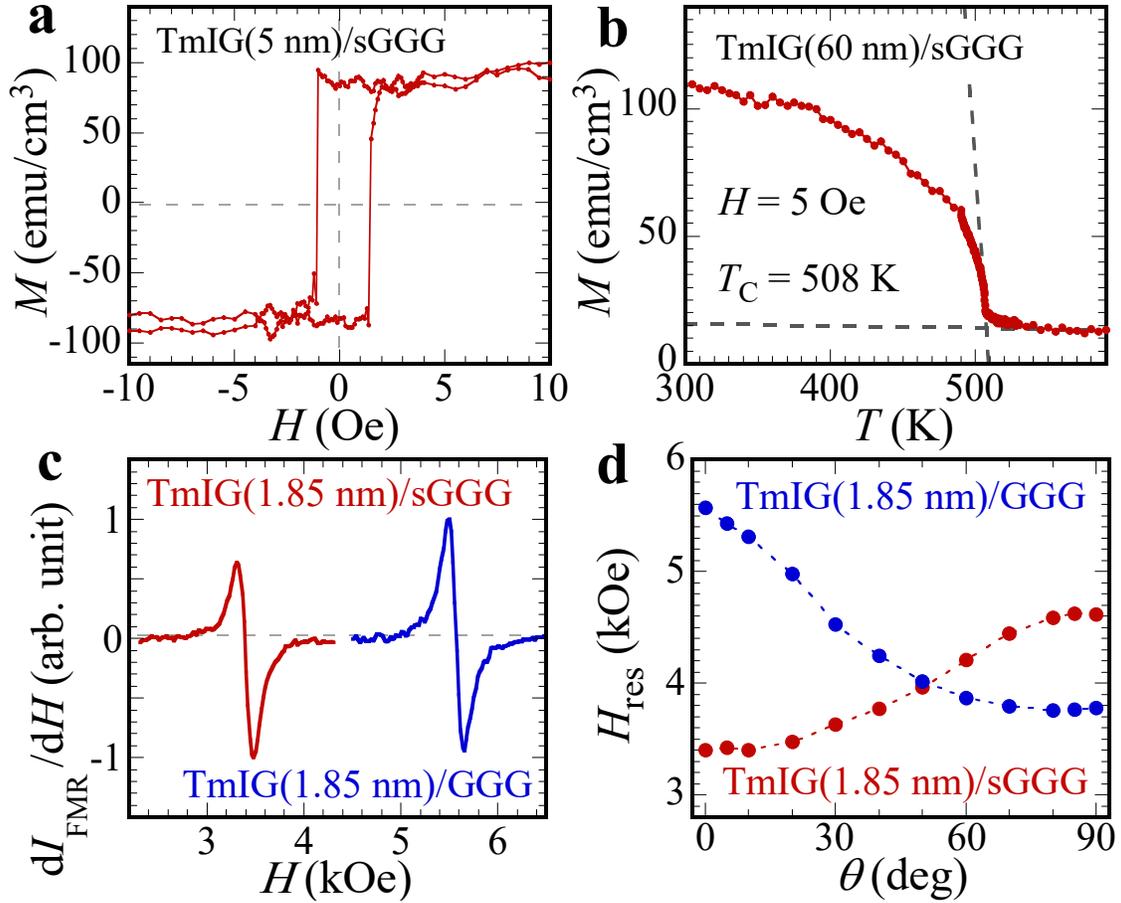

**Figure 2. Static and dynamic magnetic properties of TmIG. a**, Out-of-plane magnetic hysteresis loop of a TmIG(5 nm) film on sGGG after subtracting the linear paramagnetic substrate background. **b**, $M$ vs. $T$ plot of a TmIG(60 nm) film on sGGG with a 5 Oe field applied OOP. Dashed lines aid for the determination of $T_C$. **c**, FMR derivative spectra in an OOP field for TmIG(1.85 nm) on GGG (blue) and sGGG (red) at room temperature and 9.62 GHz. **d**, Angular dependence of FMR resonance fields for TmIG(1.85 nm)/GGG (blue) and TmIG(1.85 nm)/sGGG (red). 0° and 90° correspond to OOP and IP, respectively.

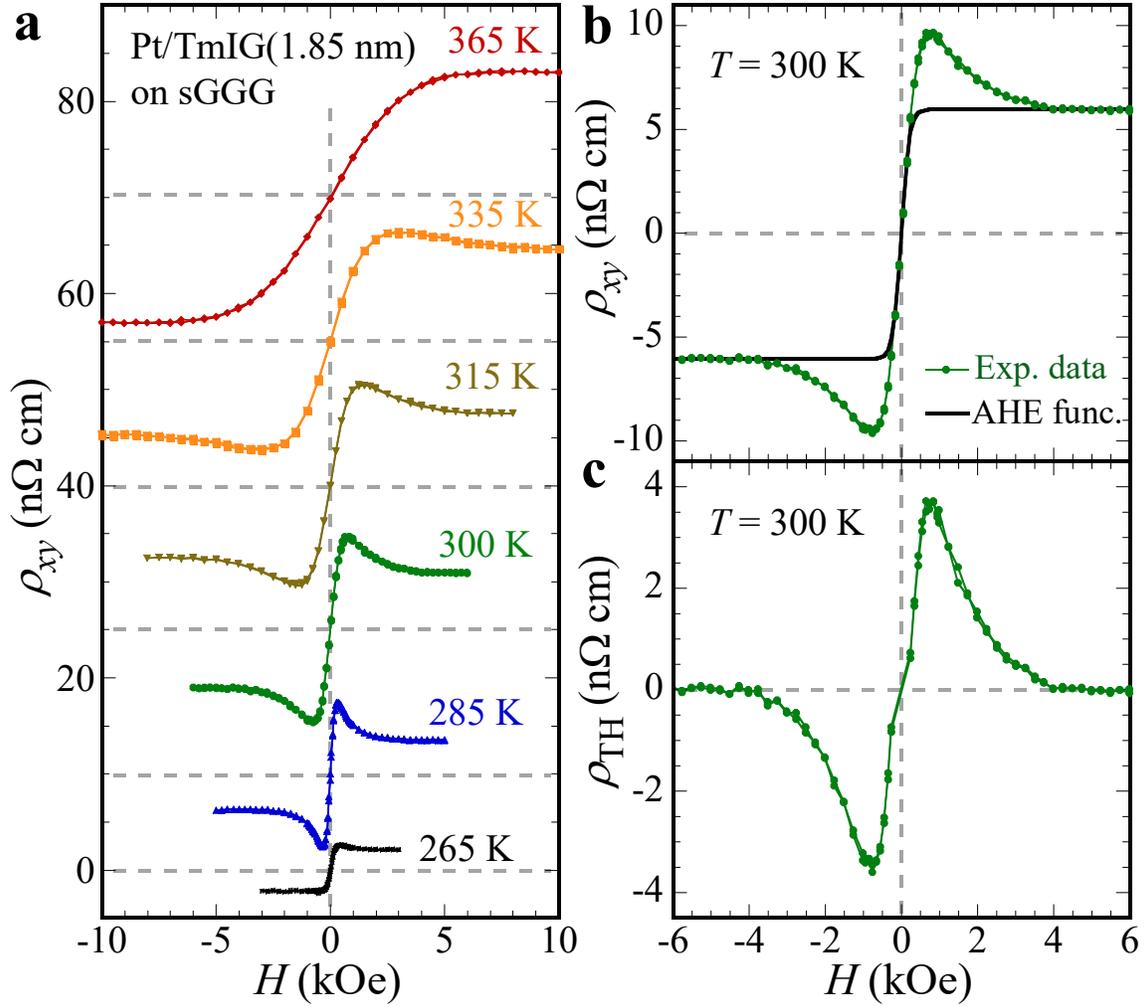

**Figure 3. Topological Hall effect in a Pt(3 nm)/TmIG(1.85 nm)/sGGG(111) bilayer. a**, Hall resistivity vs. applied magnetic field at various temperatures. The peak in $\rho_{xy}$ first appears at 265 K and vanishes at $T \geq 365$ K. Spectra are shifted vertically for clarity. **b**, Room temperature Hall resistivity where the magnetization is approximated as a $\tanh(H/H_0)$ function to represent the AHE contribution, where $H_0$ is a fitting parameter. **c**, Room temperature topological Hall resistivity generated after subtracting the two curves in **b**.

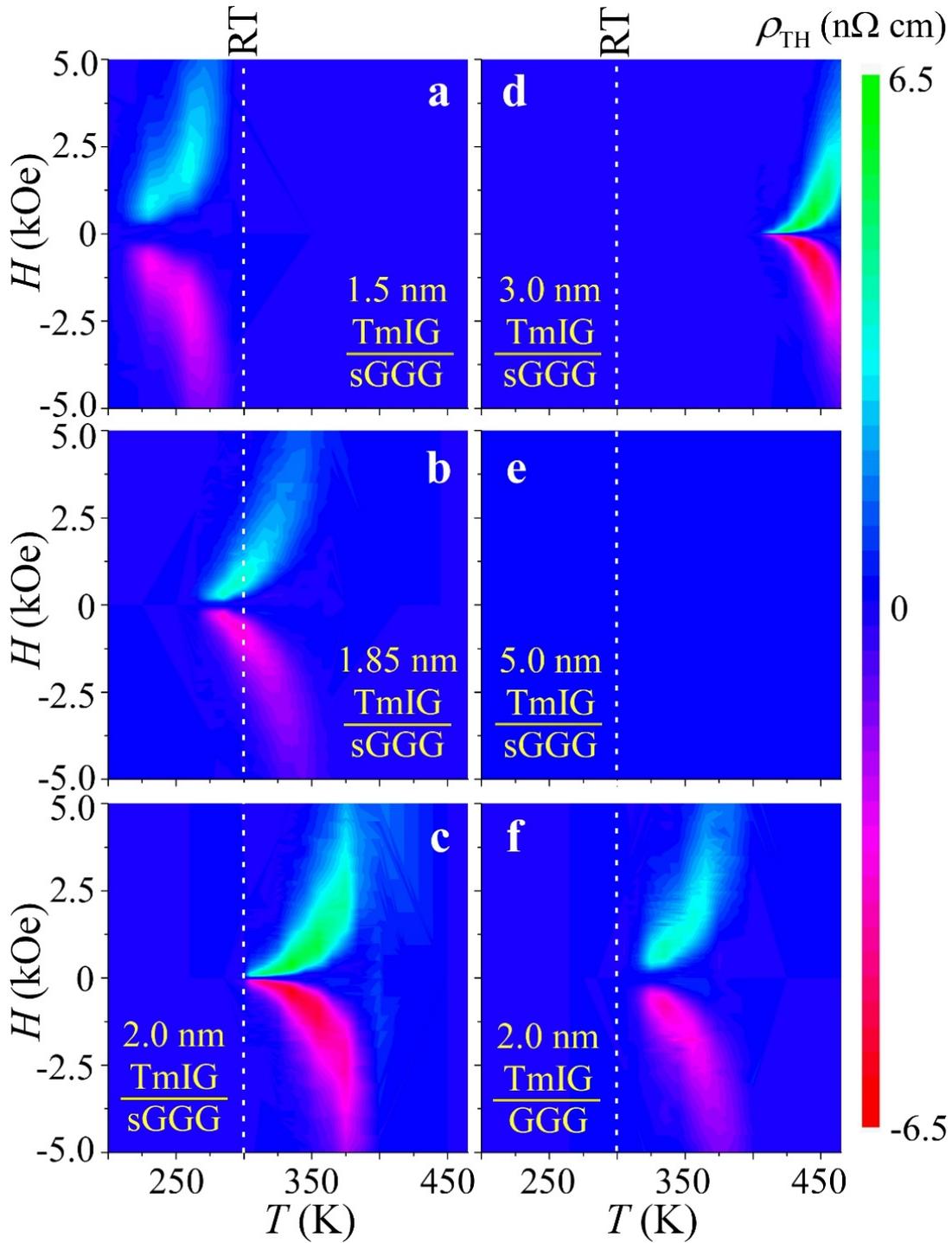

**Figure 4**. *H-T* **phase diagrams for various thicknesses of TmIG.** Topological Hall resistivity phase diagrams for Pt(3 nm)/TmIG($t$)/sGGG(111) with **a**, $t$ = 1.5 nm, **b,** $t$ = 1.85 nm, **c**, $t$ = 2 nm, **d**, $t$ = 3 nm, and **e**, $t$ = 5 nm. As the TmIG thicknesses increases, the temperature window of $\rho_{TH}$ shifts towards higher temperatures. **f**, *H-T* phase diagram for Pt(3 nm)/TmIG(2 nm)/GGG(111) where the TmIG film has an in-plane anisotropy. The vertical white dashed lines denote room temperature.